\documentclass[aps,prb,twocolumn,showpacs,superscriptaddress]{revtex4}

\usepackage{graphicx}
\usepackage{amsmath}

\begin{document}

\title{Large negative magnetoresistance in the new antiferromagnetic rare-earth dichalcogenide EuTe$_2$}

\author{Junjie Yin}
\affiliation{School of Physics, Sun Yat-Sen University, Guangzhou 510275, China }
\author{Changwei Wu}
\affiliation{School of Physics, Sun Yat-Sen University, Guangzhou 510275, China }
\author{Lisi Li}
\affiliation{School of Physics, Sun Yat-Sen University, Guangzhou 510275, China }
\author{Jia Yu}
\affiliation{School of Physics, Sun Yat-Sen University, Guangzhou 510275, China }
\author{Hualei Sun}
\affiliation{School of Physics, Sun Yat-Sen University, Guangzhou 510275, China }
\author{Bing Shen}
\affiliation{School of Physics, Sun Yat-Sen University, Guangzhou 510275, China }
\author{Benjamin A. Frandsen}
\affiliation{Department of Physics and Astronomy, Brigham Young University, Provo, UT, U.S.A.}
\author{Dao-Xin Yao}
\affiliation{School of Physics, Sun Yat-Sen University, Guangzhou 510275, China }
\author{Meng Wang}
\email{wangmeng5@mail.sysu.edu.cn}
\affiliation{School of Physics, Sun Yat-Sen University, Guangzhou 510275, China }

\begin{abstract}

We report the synthesis and characterization of a rare-earth dichalcogenide EuTe$_2$. An antiferromagnetic transition was found at T$_M$ = 11 K.  The antiferromagnetic order can be tuned by an applied magnetic field to access a first-order spin flop transition and a spin flip transition. These transitions are associated with a giant negative magnetoresistance with a value of nearly 100\%. Heat capacity measurements reveal strong electronic correlations and a reduced magnetic entropy. Furthermore, density functional theory calculations demonstrate that the electrons near the Fermi surface mainly originate from the Te 5$p$ orbitals and the magnetism is dominated by localized electrons from the Eu 4$f$ orbitals. These results suggest that both the RKKY and Kondo interactions between the local moments and itinerant electrons play crucial roles in the magnetism and large negative magnetoresistance of EuTe$_2$.
\end{abstract}

\maketitle
\section{INTRODUCTION}

Manipulation of the resistivity in magnetic or nonmagnetic metals and semiconductors via the application of a magnetic field, i.e. magnetoresistance (MR), has attracted significant interest as a probe of fundamental physics and as a phenomenon with broad potential applications. The MR effect is frequently associated with insulator-metal transitions, spin flop transitions, and spin flip transitions in magnetic materials$\cite{Dagotto2001}$. The mechanisms of the MR effect can be nontrivial and are often related to magnetic exchange interactions, Hund's couplings, and charge or orbital ordering. Notably, the giant MR (GMR) effect discovered in the magnetic multilayers of Fe/Cr in 1988$\cite{VanDau1988}$ has been widely applied in hard disks, magnetic memory chips, and many other spintronic devices.  Colossal MR (CMR) effects have been found in doped manganite perovskites, but the requirements of a high magnetic field and low temperature have limited their practical applications$\cite{Uehara1999}$. Thus, the search for new MR systems with less stringent magnetic field and temperature requirements is important.

The CuAl$_2$-type structure ($AB_2$) supports a variety of interesting physical properties and may be a good candidate for new MR materials. In this structure, each $A$ atom is surrounded by eight $B$ atoms$\cite{Havinga1972}$. The $A$ atoms in the $AB_2$ compounds are normally divalent cations such as transition metals or rare earth metals.  The $B$ atoms are generally the chalcogenides (S, Se, and Te) or antimony (Sb) and form [$B_2$]$^{2-}$ dimers stacking along the $c$ axis, as in SrS$_2$$\cite{Kawada1976}$, BaTe$_2$$\cite{Li1994}$, EuSe$_2$$\cite{Aitken2002}$, and CrSb$_2$$\cite{Sales2012}$. Properties found in materials with this structure include superconductivity in the alloy CuAl$_{2.06}$$\cite{Havinga1972a}$, thermoelectricity in Pb(Se, S)$_2$\cite{Ni2019}, and multiple magnetic transitions in EuSe$_2$$\cite{Aitken2002}$. In the latter compound, the large magnetic moments of Eu$^{2+}$ (spin $S=7/2$) may have a strong influence on the electronic properties. However, detailed studies on the magnetic and electronic properties of the Eu-based $AB_2$ system are still lacking.

 \begin{figure}[t]
\includegraphics[scale=0.35]{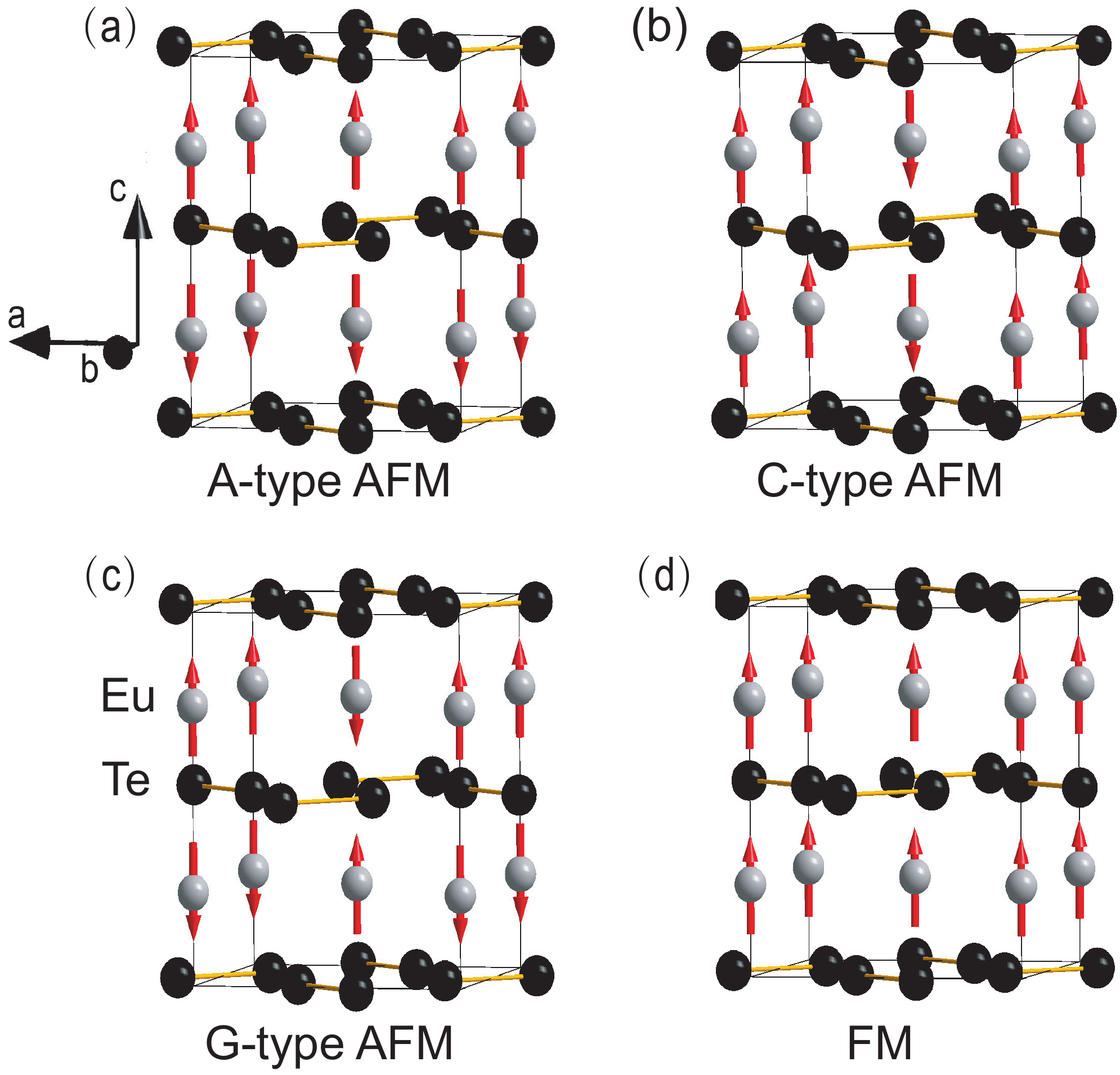}
\caption{Illustration of the crystal structure of EuTe$_2$ with four types of possible magnetic orders: (a) $A$-type antiferromagnetic order (AFM), (b) $C$-type AFM, (c) $G$-type AFM, and (d) ferromagnetic order. Eu atoms are white and Te atoms are black. The $A$-type AFM order has the lowest free energy.}
\label{fig1}
\end{figure}

\begin{figure*}
\includegraphics[scale=0.7]{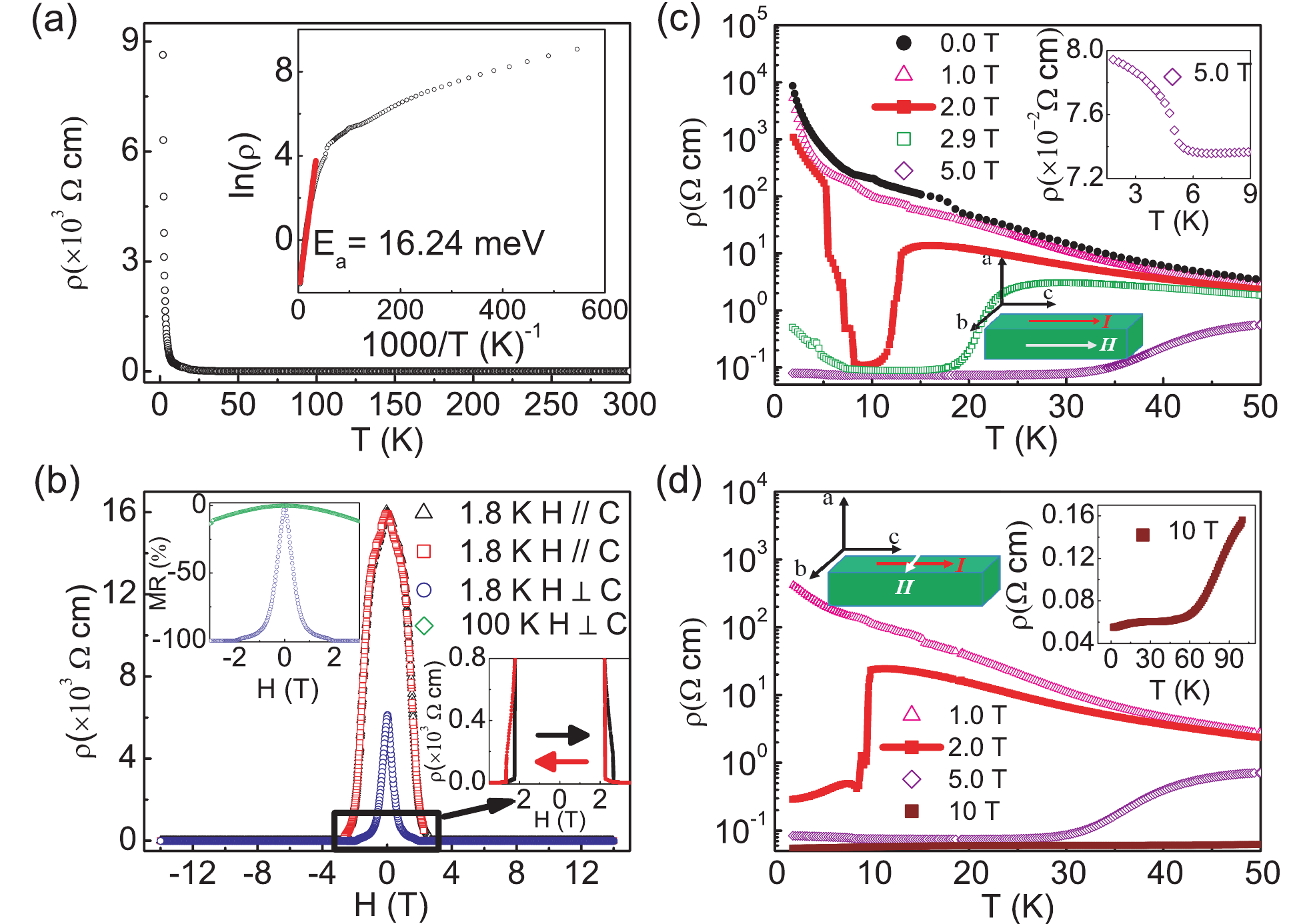}
\caption{(a) Temperature dependence of the resistivity in zero applied magnetic field. The red line in the inset shows a fit using the activation-energy-model $\rho$(\emph{T})= $\rho$$_0$ exp(\emph{E$_a$}/\emph{k$_B$}\emph{T}) (see main text for details).  (b) Magnetic field dependence of the resistivity at $T = $1.8 K with the field in the $ab$-plane and the $c$ axis. The data marked with black triangles were collected scanning from negative to positive field, while the reverse order was used for the data marked with red squares.  The inset on the right shows hysteresis of the resistivity upon application and reversal of the magnetic field. The inset on the left shows the magnetic field dependence of the resistivity at 1.8 K and 100 K ($MR = (\rho(H)-\rho(0)) / \rho(0)$) with the field perpendicular to the $c$ axis. (c) Resistivity as a function of temperature under magnetic fields of 0, 1, 2, 2.9, and 5 T parallel to the $c$ axis. (d) Identical measurements as (c) using magnetic fields of 1, 2, 5, and 10 T parallel to the $b$ axis. The insets in (c) and (d) show a closer view of the transitions at low temperatures. }
\label{fig2}
\end{figure*}

In this paper, we report the synthesis and characterization of EuTe$_2$, a new CuAl$_2$-type compound.By combining data from magnetic susceptibility, resistivity, heat capacity, and density functional theory (DFT) calculations, we determine that EuTe$_2$ is a semiconductor with a small thermal activation gap. The magnetic moments on the Eu$^{2+}$ cations order antiferromagnetically below T$_M$=11 K. This magnetic order can be manipulated by an applied magnetic field, revealing a large negative magnetoresistance attributed to the strong interactions between the local moments of Eu$^{2+}$ 4$f$ electrons and the itinerant 5$p$ electrons of the  [Te$_2$]$^{2-}$ dimers. The observed MR in EuTe$_2$ can be understood through a combination of the Ruderman-Kittel-Kasuya-Yosida (RKKY) interaction and Kondo effect.

\section{EXPERIMENT}

Our measurements were performed on single crystal samples  of EuTe$_2$. The single crystals were grown by the self flux method, similar to EuTe$_4$$\cite{Wu2019}$. Eu (99.9\%) and Te shots (99.999\%) were combined in the molar ratio $1:10$ and sealed in an evacuated quartz ampoule. The ampoule was slowly heated to 850$^{\circ}$C in 100 hours and held for 3 days, then slowly cooled to 450$^{\circ}$C in 300 hours followed by centrifuging at this temperature to separate crystals from the Te flux.  Dark and shiny crystals were obtained and used in our experiments. Energy-dispersive x-ray spectroscopy (EDX) from EVO, ZEISS was employed to determine the composition of the crystals. Single crystal X-ray diffraction (XRD) measurements were conducted on a SuperNova (Rigaku) single crystal X-ray diffractometer. The resistivity was measured using the standard four-probe method on single crystals with a typical size of $4\times1.5\times1$ mm$^3$ with a physical property measurement system (PPMS) from Quantum Design. The magnetic susceptibility and heat capacity measurements were conducted using this PPMS as well. The Vienna $ab$ initio simulation package (VASP) was employed for the DFT calculations.

\begin{table}[t]
\caption{Single crystal refinement of EuTe$_2$.}
\begin{tabular}{cc}
\hline \hline
Empirical Formula                          & EuTe$_2$     \\ \hline
EDX formula                                & Eu$_1$Te$_{2.03}$    \\
Formula weight                             & 407.164      \\
Temperature                                & 150 K   \\
Crystal system                             & Tetragonal     \\
Space group                                & I4/mcm           \\
Unit-cell parameters                       & a= b=6.9711(3) \AA  \\
                                           & c= 8.1800(5) \AA   \\
                                           &  $\alpha=\beta=\gamma$=90$^{\circ}$   \\
Atomic parameters                        & \\
Eu                                         & 4a (1/2,1/2,1/4) \\
Te                                         & 8h (x,y,1/2)\\
                                           & x = 0.6407(5), y = 0.8593(5)\\
Volume Z                                   & 397.52(3) \AA$^3$  4\\
Density                                    &8.856 g/cm$^3$ \\
Absorp. coeff.                               &29.906 mm$^{-1}$\\
F(000)                                     &688\\
Crystal size(mm)                           &0.15 $\times$ 0.12 $\times$ 0.02 \\
Crystal color                              &black\\
Radiation                                  &Mo K$_a$ ($\lambda$= 0.71073 \AA)\\
2$\Theta$ for data collection                & 4.134$^{\circ}$ to 41.899$^{\circ}$\\
Index ranges                               & -13 $\leq$ h $\leq$ 9,-9 $\leq$ k $\leq$ 12,\\
                                                        &-6 $\leq$ l $\leq$ 15\\
Reflections collected                      &1330\\
Independent reflections                    &389\\
Data/restrains/parameters                  &389/0/7\\
Goodness-of-fit on F$^2$                   &1.021\\
Final R indexes[I$\geq$2$\sigma$(I)]       &R$_1$=0.0360, wR$_2$=0.1210\\
Final R indexes(all data)                  &R$_1$=0.0375, wR$_2$=0.1233\\
Large diff.peak/hole/e \AA$^3$        &3.32/-4.82      \\ \hline \hline
\end{tabular}
\label{table:t1}
\end{table}

\section{RESULTS}
\subsection{Structure refinement }

The average composition of our single crystals was determined by EDX to be EuTe$_{2.03}$, very closed to a stoichiometric ratio of $1:2$. We took several pieces of small single crystals for x-ray diffraction measurements. The refined results are consistent with a CuAl$_2$-type structure, with the key results summarized in Table \ref{table:t1}. The space group is tetragonal $I4/mcm$ at 150 K, similar to its sister compound EuSe$_2$\cite{Aitken2002}. The nearest neighbor Eu$^{2+}$ atoms are located along the $c$ axis with a separation distance of 4.09 \AA. The Te atoms forms [Te$_2$]$^{2-}$ dimers, which stack along the $c$ axis. The $c$ axis is somewhat elongated compared to the identical $a$ and $b$ axes. The crystal structure is illustrated in Fig. \ref{fig1}, together with various potential magnetic orders that will be discussed subsequently. 

\subsection{Resistivity}

A voltage was applied along the $c$ axis for all the resistivity measurements, which are summarized in Fig. \ref{fig2}. The data in Fig. \ref{fig2} (a) reveal clear semiconducting behavior. A fit of the thermal activation gap was performed on the data from 30 to 300 K using the activation-energy model $\rho$(\emph{T})= $\rho$$_0$ exp(\emph{E$_a$}/\emph{k$_B$}\emph{T}), where  $\rho$$_0$ is a prefactor and \emph{k$_B$} is the Boltzmann constant. The refined activation energy is $E_a=16.24$ meV.  The resistivity has a value of 8.6$\times$10$^3$ $\Omega$ cm at 1.8 K and decreases to 1.3$\times$10$^{-1}$ $\Omega$ cm at room temperature. The sharp upturn at low temperature is consistent with the hybridization of electrons due to the formation of magnetic order$\cite{Kondo2005,Kamihara2008,Wu2016,Cheng2019}$.

To explore the influence of an applied magnetic field on the resistivity, we measured the resistivity at 1.8 K under a field up to 14 T. A giant negative MR was observed for field applied both parallel and perpendicular to the $c$ axis, as shown in Fig. \ref{fig2} (b). The giant negative MR disappears at 100 K (left inset of Fig. \ref{fig2} (b)). Scanning the field from 14 T to -14 T and back in the $H // c$ configuration results in hysteresis of the resistivity (right inset of Fig. \ref{fig2}(b)), evidencing a field-induced first-order transition associated with magnetism in this material. A similar phenomenon has been observed in Nd$_{0.5}$Sr$_{0.5}$MnO$_3$$\cite{Kuwa1995}$.

The temperature dependence of the negative MR was investigated for fields of 0, 1, 2, 5, and 10 T applied along the $c$ axis and along the $a$ or $b$ axis (they are equivalent due to the tetragonal crystal structure). At 1.8 K, the resistivity decreases by 5 orders of magnitude  under a field of 5 T compared to that with no applied field. This semiconductor-to-metal transition moves to a higher temperature and becomes broader under larger magnetic fields$\cite{Yoshinori1994,Felsch1973,Malick2019}$. Interestingly, the resistivity shows an upturn at low temperatures, indicating a re-entry to the semiconducting state. This is shown in Fig. \ref{fig2} (c). The semiconductor-to-metal and metal-to-semiconductor transition temperatures extracted from the resistivity measurements under magnetic fields have been summarized in Fig. \ref{fig6}. As seen in Fig. \ref{fig2} (d), the re-entry of the semiconducting state at low temperatures when the field is applied along the $b$ axis is not obvious at 2 T. However, the upturn of resistivity still exists below 8.4 K for 5 T. For the field of 10 T, the upturn of resistivity disappears as shown in the inset of Fig. \ref{fig2} (d). 

\begin{figure*}
\centering
\includegraphics[scale=0.65]{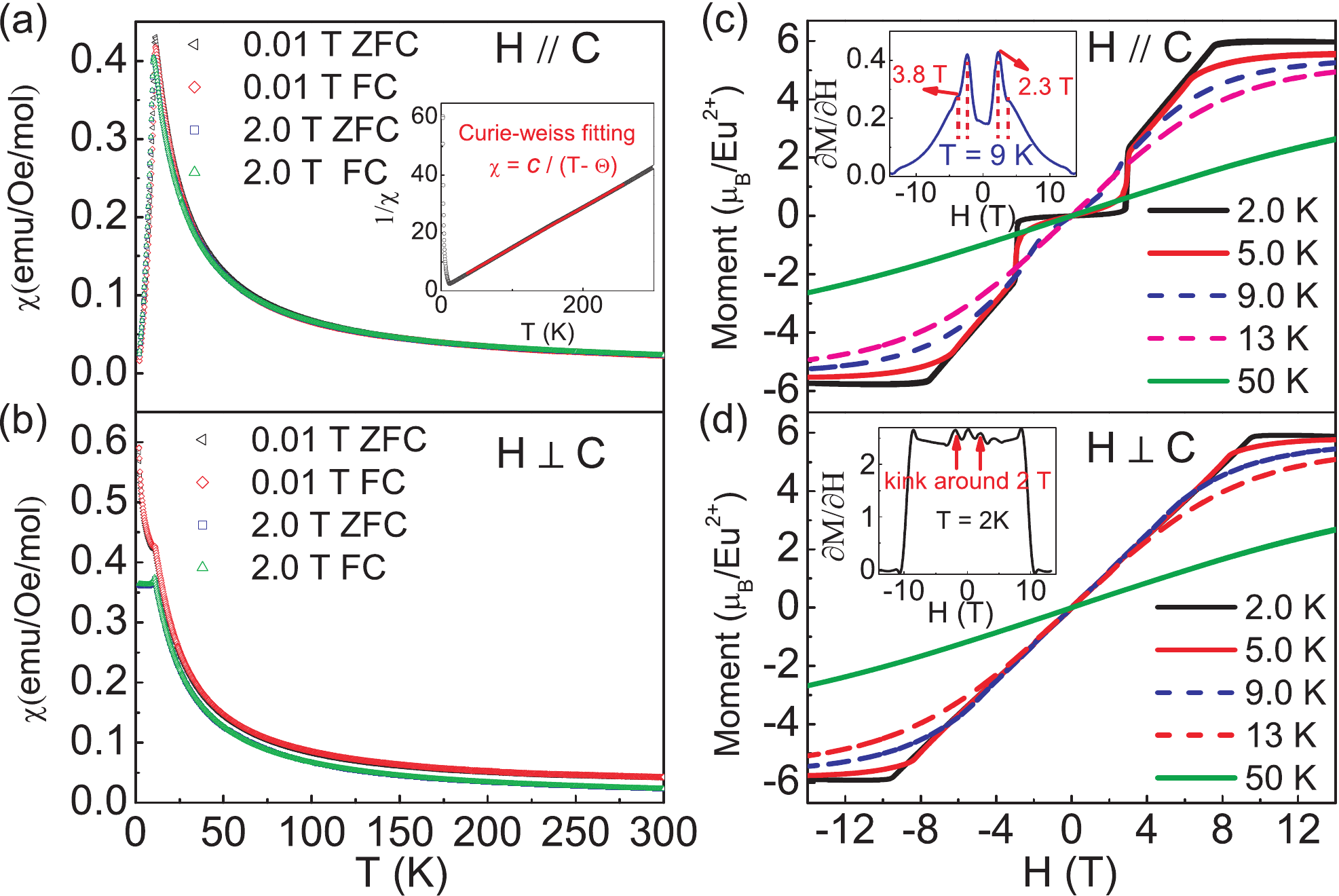}
\caption{Field cooling (FC) and zero field cooling (ZFC) measurements of the magnetic susceptibility under fields of 0.01 and 2 T (a) parallel to the $c$ axis and (b) perpendicular to the $c$ axis. The inset in (a) is a Curie-Weiss fit using the function $\chi=C/(T-\Theta)$. (c-d) Field-dependent magnetization at various temperatures with the field $H\parallel c$ (c) and $H\bot c$ (d) from 2 to 50 K. The insets in (c) and (d) are derivatives of the magnetization as a function of magnetic field.}
\label{fig3}
\end{figure*}

\subsection{Magnetic susceptibility}

Magnetic susceptibility measurements as a function of temperature displayed in Fig. \ref{fig3} (a, b) reveal an antiferromagnetic (AFM) transition at $T_M=11$ K. The Curie-Weiss law $\chi = C / (T - \theta)$ was employed to fit the susceptibility, where \emph{C} is the Curie constant and $\theta$ is the Curie-Weiss temperature. The fitting reveals an effective moment $\mu$$_{eff}$ = 7.5 $\mu$$_B$ and $\theta=-6.05$ K. The effective moment $\mu_{eff}$ is very close to the theoretical expectation of $\mu=\sqrt{S(S+1)}g\mu_B=7.9 \mu_B$ for Eu$^{2+}$, where $g=2$ is the spin Landau $g$ factor of electrons and spin $S=7/2$ is expected for Eu$^{2+}$. This negative $\theta$ agrees with the AFM transition below $T_M$. The field direction dependence of the susceptibility at low temperatures indicates that the moments of the AFM order align along the $c$ axis.

To investigate further the dependence of the susceptibility on the applied field, we performed isothermal magnetization ($M$) measurements as presented in Fig. \ref{fig3} (c, d). At 2 K, several obvious transitions could be observed in the magnetization in the case of the field applied along the $c$ axis (panel c). The abrupt increase at 3 T corresponds to a spin flop transition of the AFM order. The magnetization saturates at $M_{sat}=5.89 \mu_B$ per Eu$^{2+}$ ion for magnetic fields above 7.6 T at 2 K, marking the spin flip transition. The $M_{sat}$ moment is smaller than that of the theoretical estimation $M=gS\mu_B=7.0 \mu_B$. This could be explained by itinerant electrons contributed by Te screening the localized moments on the Eu sites. With increasing temperature, the spin flop and spin flip transitions are weakened and finally vanish above $T_M$. The  transition temperatures have been extracted and plotted in the $T-H$ phase diagram in Fig. \ref{fig6}. When the field is applied in the $ab$ plane (Fig. \ref{fig3}(d)), the spin flip transition dominates, while the spin flop transition is not readily observable. The spin flip field of 10 T for $H\bot C$ is larger than that of 7.6 T for $H\Vert C$, pointing to anisotropic magnetism in EuTe$_2$. Small kinks could be observed in the $\partial M/\partial H$ curve in Fig. \ref{fig3} (d). The kinks may be caused by the imperfect alignment of the piece of single crystal.

The transitions seen in the magnetization data are consistent with the magnetoresistance transitions in Fig. \ref{fig2}. This may suggest that the large negative MR is related to the Kondo lattice scattering by the localized antiferromagnetically ordered Eu$^{2+}$ moments, as observed in Na$_{0.85}$CoO$_2$$\cite{Luo2004}$, Ca$_{1-x}$Sr$_x$Co$_2$As$_2$$\cite{Ying2012}$, and Ca$_{2-x}$Sr$_x$RuO$_4$$\cite{Nakatsuji2003}$.

\begin{figure}
\includegraphics[scale=0.3]{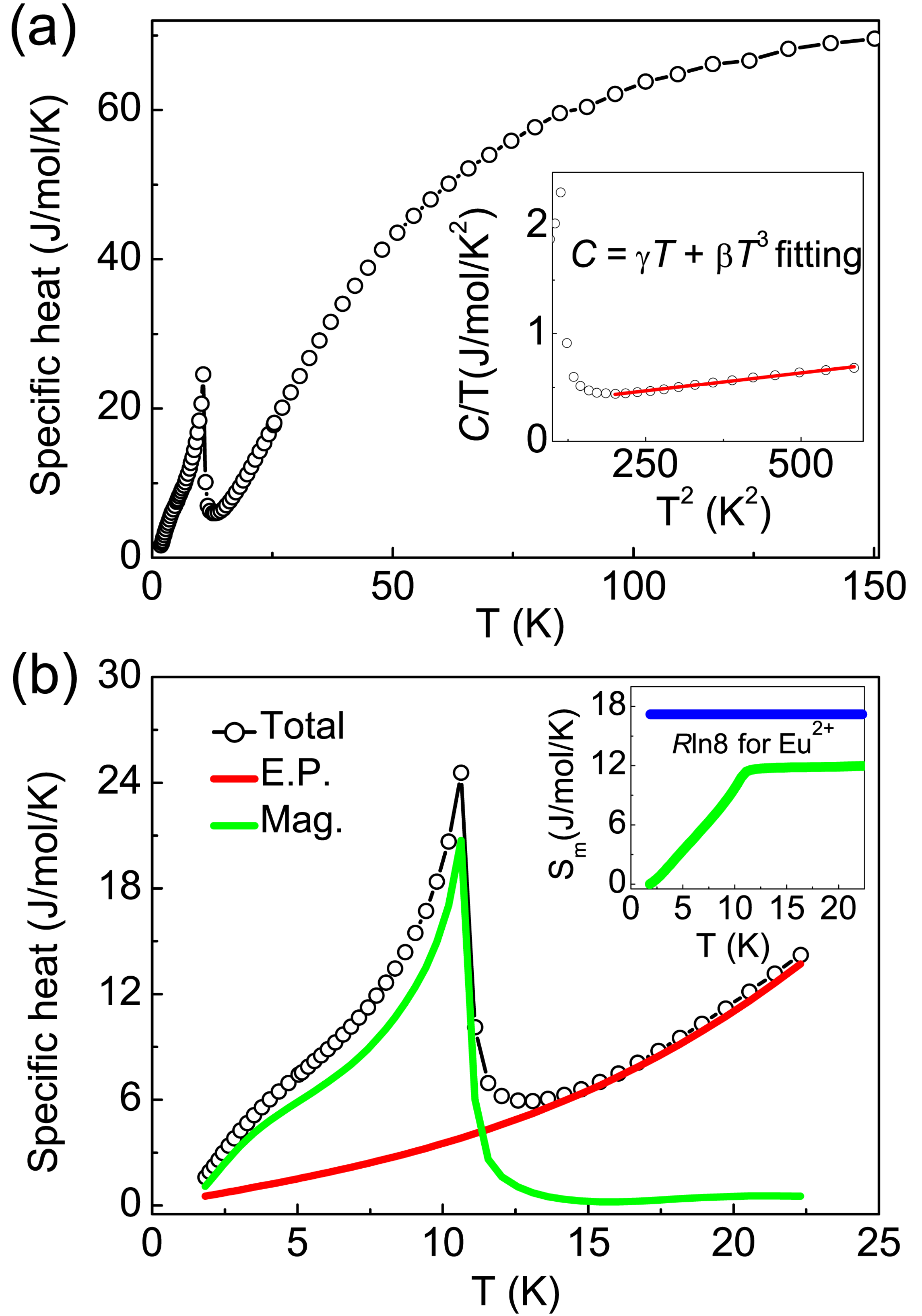}
\caption{(a) Heat capacity of EuTe$_2$ as a function of temperature. The inset shows a fit to the heat capacity using the Fermi liquid model $C=\gamma T+\beta T^3$. (b) Separation of the electronic and phonon (E.P.) and magnetic (Mag.) contributions to the heat capacity by extrapolating the fit in (a). The inset shows the magnetic entropy of Eu$^{2+}$ and a theoretical value ($Rln8$) for Eu$^{2+}$.}
\label{fig4}
\end{figure}

 \begin{figure}
\includegraphics[scale=0.2]{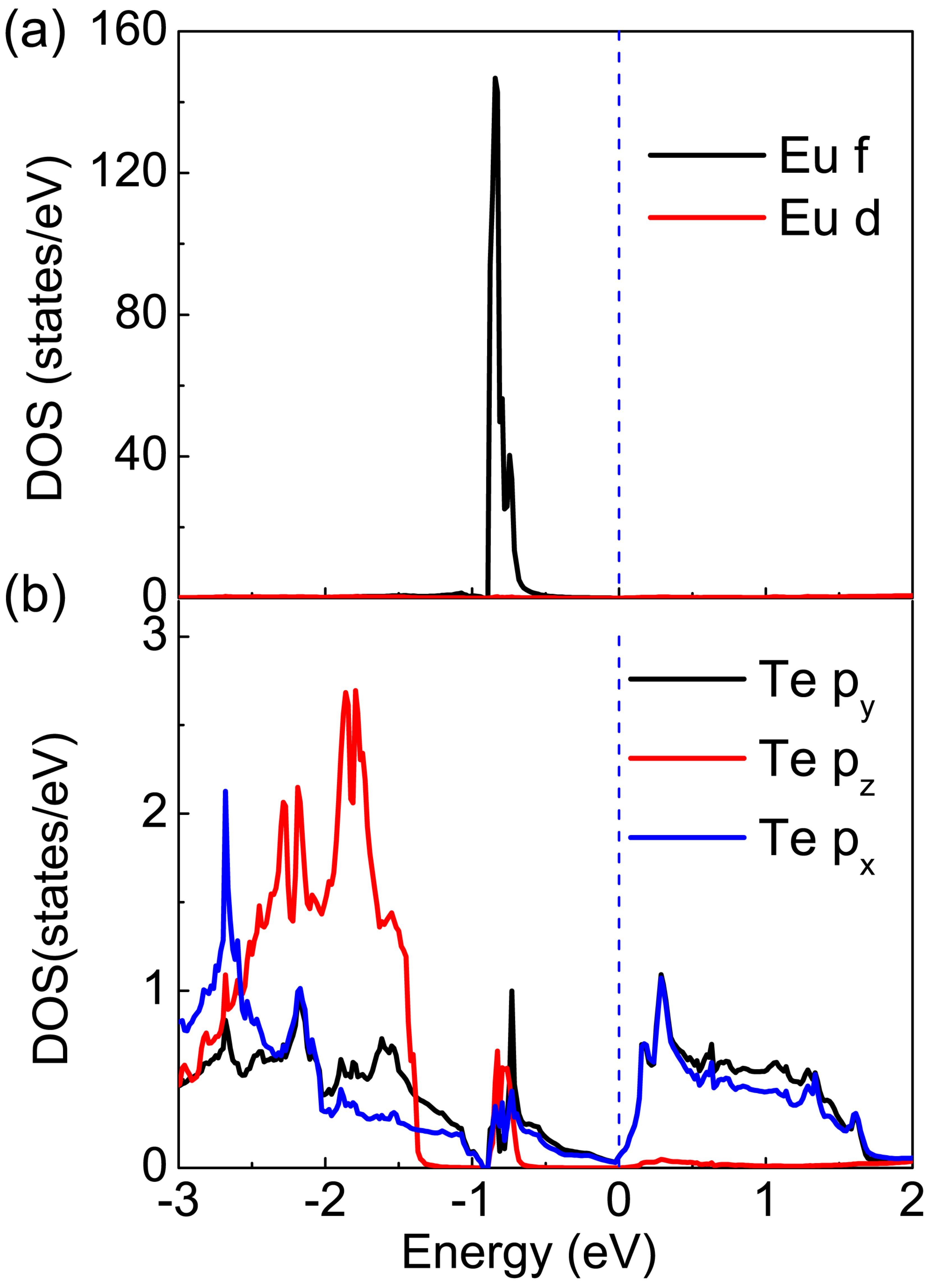}
\caption{Density functional theory (DFT) calculations of (a) the density of states (DOS) of Eu$^{2+}$ 4$f$ and (b) Te$^{-}$ 5$p$ electrons in EuTe$_2$.  The valence of $1-$ is an average of the Te$_2^{2-}$ dimer.}
\label{fig5}
\end{figure}

\subsection{Specific heat capacity}

In Fig. \ref{fig4} (a), we show the heat capacity measurements. A sharp $\lambda$-like transition occurs at $T_M$ = 11 K, which is consistent with the AFM transition observed in the magnetic susceptibility measurements shown in Fig. \ref{fig3} (a, b). A Fermi-liquid model $C=\gamma T + \beta T^3$ is used to fit the specific heat capacity at temperatures above  $T_M$, resulting in $\gamma$ = 303.7 mJ/mol/K$^2$ and $\beta$ = 0.666 mJ/mol/K$^2$. The first term $\gamma T$ represents the contribution of electrons and the second term $\beta T^3$ represents phonons. The large electronic coefficient $\gamma$ indicates the presence of strong electronic correlations.
Furthermore, we can obtain the Debye temperature $\theta$$_D$ = [12$\pi$$^4$NR/(5$\beta$)]$^{1/3}$ = 206 K for EuTe$_2$, where $N$ is the number of atoms in the chemical formula, and $R=8.314$ J/mol/K is the ideal gas constant.

To obtain the magnetic contribution to the heat capacity, we extrapolated and subtracted the electron and phonon contributions at low temperatures, as shown in Fig. \ref{fig4} (b). The magnetic entropy S$_{mag}$ = $\int_{0}^{T}$\emph{C}$_{mag}$/\emph{T} \emph{d}\emph{T} is calculated to be 11.46 J/mol/K, which is smaller than that of the theoretical expectation with \emph{S}$_m$ = \emph{R}$\ln$(2\emph{S}+1)= 17.29 J/mol/K for Eu$^{2+}$. The integration range $T=20$ K was used in our calculation. Reduced magnetic entropy has been reported in cerium-based compounds$\cite{Hegger2000,Chen2017}$, which was attributed to the competition between the RKKY interaction and Kondo interaction. In other words, the localized magnetic moments are partially screened by the itinerant electrons through the Kondo interactions$\cite{Yang2008,Yang2017,Nakatsuji2004}$.

\subsection{DFT calculations}

To investigate the magnetic structure of EuTe$_2$, DFT calculations have been performed to compare the free energies of four magnetic structures,  as illustrated in Fig. \ref{fig1}. The calculated free energies for $A$-type, $C$-type, $G$-type, and ferromagnetic (FM) orders are -58.5270 eV, -58.5209 eV, -58.4557 eV, and -58.1396 eV, respectively. The $A$-type AFM order therefore has the lowest predicted energy, suggesting it is most likely the magnetic ground state for EuTe$_2$.

In addition, we calculated the electronic density of states with $A$-type AFM order. To account for the strong correlation effects, the on-site Coulomb interaction $U$ was introduced. The impact of $U$ on the magnetic ground state was investigated using values of $U$ ranging from 0 to 7 eV. The experimentally observed magnitude of the Eu$^{2+}$ ordered moment is reproduced with $U=6\sim7$ eV. Hence, we used $U=6$ eV for Eu$^{2+}$ in our subsequent calculations. The results are summarized in Fig. \ref{fig6}, indicating that localized Eu 4$f$ electrons reside 0.9 eV below the Fermi level and support the localized magnetic moments, while the Te 5$p$ orbitals are more spread out in energy and make a weak contribution to the density of states at the Fermi level. The hybridization of the Eu 4$f$ Te 5$p$ electrons may result in the observed semiconducting state of EuTe$_2$ at low temperatures$\cite{Mott1974}$.

\section{Discussion and summary}

\begin{figure}
\includegraphics[scale=0.4]{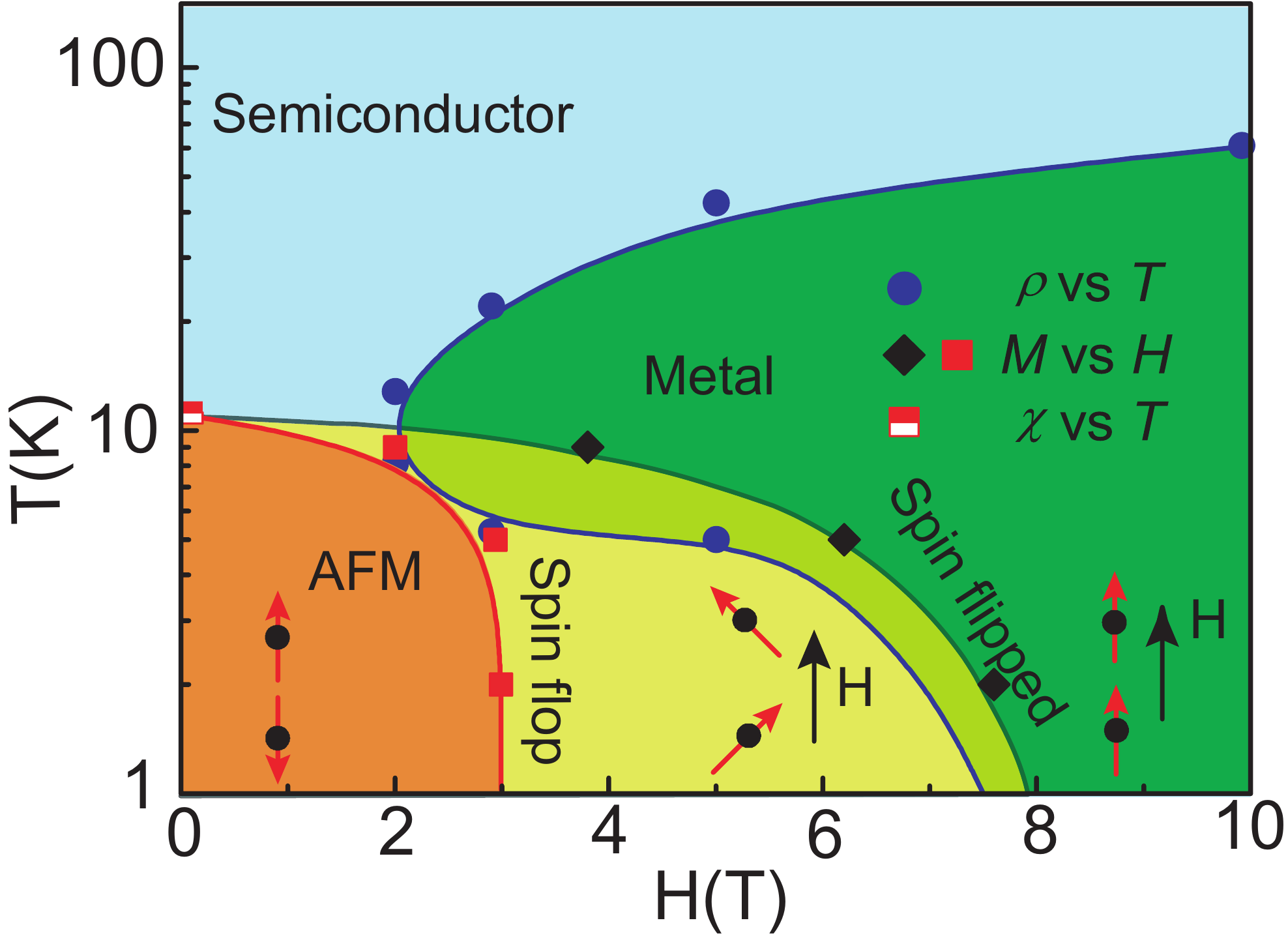}
\caption{A proposed $T-H$ phase diagram for EuTe$_2$. The sample undergoes a transition from a semiconducting state to a metallic state by applying a magnetic field. The metallic region is shaded green.  Below the AFM transition temperature $T_M$, the spins flop at around 3 T and flip progressively as the magnetic field increases until reaching saturation as marked by the black diamonds. }
\label{fig6}
\end{figure}

The results presented here reveal rich temperature- and field-dependent phase behavior in EuTe$_2$. This is summarized by the $T-H$ phase diagram in Fig. \ref{fig6}. The semiconductor-to-metal transition temperatures were extracted from the resistivity measurements, while the antiferromagnetic, spin flop, and spin flip transitions were extracted from the magnetic susceptibility measurements with the field applied parallel to the $c$ axis (and therefore also parallel to the ordered moments). The whole $T-H$ phase diagram can be divided into a semiconducting region and a metallic region based on conductance. Below $T_M=11$ K, we observe an AFM region, a spin flop transition region showing field-induced hysteresis, and a saturated spin flip region. The details of these magnetic states and transitions are governed by the relative coupling strengths of the antiferromagnetic exchange interaction, easy axis anisotropy, and uniaxial single-ion anisotropy to the magnetic field, consistent with the expectations from mean-field theory\cite{Li2016}. The re-entrant semiconducting region for fields around $2\sim8$ T largely coincides  with the spin flip transition region, suggesting the existence of RKKY interactions between the localized 4$f$ electrons of Eu and the itinerant 5$p$ electrons of Te. However, the formation of the electronic gap, the large electronic coefficient $\gamma$, the reduced saturated magnetization, and the reduced magnetic entropy can be naturally understood in the Kondo lattice scattering scenario, in which hybridization and screening effects exist at low temperatures. Thus, our data demonstrate that both of these competing interactions play a role in the negative MR and magnetic transitions of EuTe$_2$.

The behavior of the resistivity with the coupling to the magnetism could be interpreted by the competitions between the thermal fluctuations, magnetic exchange interactions, easy axis anisotropy, and the magnetic torques of magnetic field to the local moments. In addition, both the spin-up and spin-down conducting electrons are scattered by the antiferromagnetically ordering moments of Eu$^{2+}$, as a result, the resistivity would be increased rapidly when the system enters into the AFM state. While the applied magnetic field suppresses the AFM order and drives the local moments to the spin flipped state, the magnetic scatterings between the polarized moments and the electrons with the identical spin polarization are intensively weaken. Thus, the resistance becomes much smaller. However, the giant negative magnetoresistance is merely observed in bulk compounds.  The value of the negative MR nearly $-100\%$ at 2 K and $H=2$ T is larger than that of the TiTeI nanosheets ($-85\%$)$\cite{Guo2014}$, Fe/Cr/Fe heterostructure ($-60\%$)$\cite{VanDau1988}$ and the perovskite-like La-Ba-Mn-O magnetic thin film ($-60\%$)$\cite{VonHelmolt1993}$.

In summary, we have successfully synthesized and characterized a new rare-earth dichalcogenide system EuTe$_2$. By combining structural refinements, resistivity, magnetic susceptibility, specific heat capacity, and DFT calculations, we demonstrate that EuTe$_2$ exhibits antiferromagnetic order most likely of the $A$ type which can be easily tuned with a magnetic field. Giant negative magnetoresistance that couples with the magnetic states is observed.  Both the RKKY interaction and the Kondo lattice effect play a role in the negative MR and magnetic transitions. Our work opens up new prospects to look for negative magnetoresistance in Eu-based magnetic materials.

M. W. thanks fruitful discussions with Shiliang Li and Haifeng Li. The research was supported by NSFC-11904414 and NSF of Guangdong under Contract No. 2018A030313055, the Hundreds of Talents program of Sun Yat-Sen University, and Young Zhujiang Scholar program. H. L. Sun was supported by NSFC-11904416. C.W.  and D.X.Y. acknowledge support from NKRDPC-2018YFA0306001, NKRDPC-2017YFA0206203,  NSFC-11574404, National Supercomputer Center in Guangzhou, and Leading Talent Program
of Guangdong Special Projects.



\bibliography{EuTe2}

\end{document}